\documentstyle[twocolumn,aps,epsf]{revtex}

\begin{document}

\twocolumn[
\hsize\textwidth\columnwidth\hsize\csname@twocolumnfalse\endcsname
\draft

\title{Internal avalanches in a pile of superconducting vortices}
\author{Kamran Behnia and Cigdem Capan}
\address{Laboratoire de Physique des Solides (CNRS), Universit\'e Paris-Sud, 
91405 Orsay, France }
\author{Dominique Mailly and Bernard Etienne}
\address{Laboratoire de Micro\'electronique et de Microstructures (CNRS), 196
 Ave. Ravera, 92220 Bagneux, France}
\date{\today}

\maketitle

\begin{abstract}
Using an array of miniature Hall probes, we monitored the spatiotemporal
variation of the internal magnetic induction in a superconducting niobium sample 
during a slow sweep of  external magnetic field. We found that a sizable fraction 
of the increase in the local vortex  population occurs in abrupt jumps. The size 
distribution of these avalanches presents a power-law collapse on a limited
range. In contrast, at low temperatures and low fields, huge
avalanches with a typical size occur and the system does not display a well-defined 
macroscopic critical current.    
\end{abstract}

]

Magnetic field penetrates type II superconductors as a population of
quantized flux lines called vortices due to the associated currents whirling
around them. In the absence of disorder, the repulsive force between vortices
leads to the formation of a regular lattice. Vortex movement as a result of
the Lorentz force produced by an applied current produces dissipation and
destroys superconductivity. Thus, it has a significant technological
importance. The interaction between the vortex lattice and the crystal
defects which pin vortices down creates a wealth of physical phenomena\cite
{blatter} including thermodynamic phase transitions\cite{safar}. While
current-driven movement of vortices has been intensely explored during the
past decade\cite{crabtree}, out-of-equilibrium properties of vortex matter
in absence of applied current have been subject to few studies.

As Bean showed many years ago\cite{bean}, the distribution of vortices
entering a superconducting sample is inhomogeneous. A finite gradient in
vortex density builds up to create a driving force inward balanced by the
pining forces opposing vortex movement. The slope of this pile of vortices-
originally compared by de Gennes to a sandpile\cite{degennes}- is
proportional to the local magnitude of the critical current. Recently, the
dynamic properties of such a pile has attracted new attention\cite
{field,olson,bassler}. This interest has emerged in the context of the
introduction of the concept of self-organized criticality(SOC) a decade ago 
\cite{bak} and covers\ a number of different marginally stable systems
assumed to present a comparable dynamic response when driven to the threshold of
instability. Apart sandpiles\cite{nagel,frette} which played the role of a
paradigm in the field, the list includes magnetic domains presenting \
Barkhausen noise \cite{spasojevic}, microfractures\cite{zapperi} ,
earthquakes and other complex systems. In the case of a type
II superconductor in the Bean state, addition of vortices by a slow increase
in the external magnetic field -analogous to the introduction of new grains
to a sandpile- is expected to produce vortex avalanches of all sizes and
maintain a constant gradient in flux density. Such a SOC type of behavior
has been reproduced numerically \cite{olson,bassler} and the existence of
these vortex avalanches has been experimentally established by Field and
co-workers\cite{field} who recorded the voltage pulses produced by sudden
changes of the flux density in a superconducting tube during a slow ramping
of magnetic field. They detected avalanches of various magnitudes with a
distribution exhibiting a power-law collapse over one decade. However,
their experimental set-up only resolved avalanches occurring due to sudden
exit of flux lines. In analogy with early sandpile experiments monitoring
only grains falling off the system\cite{held}, this configuration could not
detect internal events occurring within the sample. Direct observation of
internal avalanches in sandpiles was an important experimental refinement 
\cite{bretz} which paved the way to demonstrating the limited relevance of
SOC to real granular matter\cite{frette}.

In this letter, we present a study of spatiotemporal profile of local
magnetic induction in a superconductor during a slow ramp of an external
magnetic field. By directly monitoring the population of vortices in a given area of
the sample, we found that a sizeable fraction of the increase in the flux
density occurs as abrupt jumps and the distribution of size and duration of
these avalanches present a power-law behavior on a limited range. On the
other hand, at low temperatures and fields, the same\ sample presents a
different regime associated with huge avalanches. According to our results,
the dynamic behavior of superconducting vortices is more complex than what
is expected  by recent simulations predicting SOC-type response
with universal exponents\cite{olson,bassler}.

The 0.8 $\times $ 0.8 mm$^{2}$ sample studied in our work was a granular
film of niobium 20 $\mu m$ thick. A 8$\times $8 matrix of Hall probes was
used to measure local magnetization. Each element of this matrix was a
GaAlAs/GaAs heterojunction($\mu =170m^{2}/Vs$ and n= 1.7 10$^{11}m^{-2}$ at 4 K)
monitoring the magnetic field within its area (20 $\times $ 5 $\mu m^{2}$).
In principle, this configuration yields a two-dimensional profile of the
magnetic field, but in our experiment we only processed data provided by two
array-like rows of the matrix. Using a set of Burr-Brown INA128
preamplifiers we achieved a resolution of 5 10$^{-7}$ T/$\sqrt{Hz}$ in 
measuring magnetic inductions of the order of 0.1T . High-speed acquisition was
performed using a 16-bit 100kHz A/D converter.

\begin{figure}[tbp]
\epsfxsize=8.5cm
 $$\epsffile{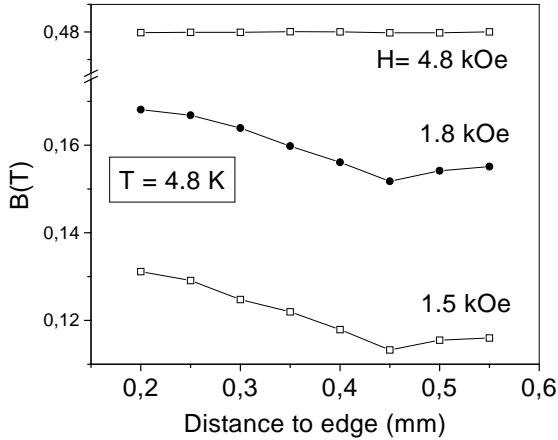}$$
\caption{The static profile of the magnetic induction in the sample at
various fields as seen by the array of Hall probes.}
\end{figure}
We thoroughly studied the magnetization features of our sample which proved
to present strong pinning features associated with a hysteretic
magnetization below T$_{c}$ (9.1 K). Moreover, the direct measurement of the
spatial variation of the magnetic field permitted to establish the presence
of a Bean-type field profile in the sample. As seen in Fig. 1, the penetration
of magnetic field in the sample is not uniform and the magnetic induction
decreases almost linearly inward. Pinning forces opposing the uniform distribution 
of vortices create a field gradient (8 10$^{-3}$ T m$^{-1}$
at T= 4.8 K and H= 1500 Oe). The critical current is thus estimated to be 
10$^{8}$ A m$^{-2}$).

To explore the criticality of this vortexpile, we studied the
time-dependence of the local field in response to a gradual increase in the
external magnetic field. Fig. 2 presents the local magnetic induction as
monitored by one Hall sensor during a slow ramp (1.1 Oe/s) of the external
field beginning at H = 1500 Oe. The temperature was kept constant at 4.8K.
As seen in the figure, the magnetic induction- that is, the vortex
population- increases steadily during this ramp. However, by magnifying
portions of the B(t) curve, it is readily observed that this increase is
not smooth and contains many abrupt jumps of many different scales. By
converting the magnitude of local induction to flux density, one finds that
during the 128 s duration of the field ramp, 1125 new vortices enter to the
area covered by the Hall sensor which already contained about 6750 vortices.
Our data shows that the addition of these new vortices leads to frequent
sudden rearrangements. We checked that no such abrupt jumps occur in the
normal state (for temperatures higher than T$_{c}$ or for fields exceeding 
H$_{c2}$). In analogy with sandpiles, the term avalanche is used to designate
these abrupt changes in vortex arrangement. Note that--- contrary to
periodic magnetothermal instabilities observed at very fast field ramps\cite
{monier}--- these avalanches occur aperiodically at millisecond timescales
and are not related with any temperature instability of the sample. In order 
to determine the statistical distribution of avalanche size and
durations, we analyzed the data from four field ramps starting at H=0.15T
identical to the one presented in Fig.2. Our sensitivity permitted to
resolve abrupt jumps as small as 0.08 G ( 0.16 $\Phi _{0}$). We found some
variation in the ratio of avalanche activity in the output of different
sensors presumably due to inhomogeneous distribution of defects in the
sample. Typically, between 40 to 70 percent of the increase in the vortex
population occurs by such detectable jump. We recall that in the pioneer
experiment of Field and co-workers\cite{field} which was sensitive to
avalanches containing at least 50 vortices, only 3 \% of the increase in the
vortex population was resolved as detectable avalanches. Hence, our work, by
lowering the proportion of the increase in flux density which may be
attributed to fluidlike movement of the vortices, highlights the granular
nature of the vortex matter.

\begin{figure}[tbp]
\epsfxsize=8.5cm
 $$\epsffile{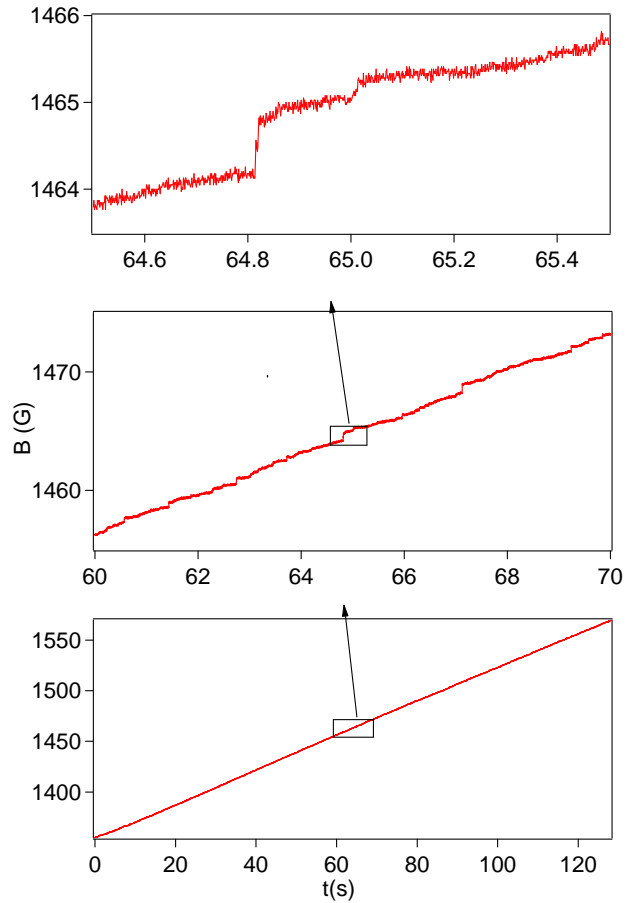}$$
\caption{The local field recorded by the first Hall captor as a function of
time. The external field was ramped at a rate of 1.1 Oe/s starting at H=
1500 Oe. Upper frames are magnifications of selected areas of lower frames.}
\label{Fig. 2}
\end{figure}

Solid circles in Fig. 3 represent the size distribution of these avalanches.
A power-law distribution P(s)= s$^{-\tau }$ with an exponent $\tau $=-2.1
fits the size distribution of smaller avalanches. This exponent should be
compared with those reported by the previous experiment(-1.4 to -2.2\cite
{field}) and simulations (-1.6\cite{bassler}). However, the range of
validity for this behavior is less than one decade. Its lower bound meets
the instrumental noise (0.08 G). The size of the largest avalanche detected
was about 1.1G which is associated with the sudden entry of 5 vortices to
the detection area. However as seen in the figure, there is a downward
deviation from the power-law behavior for avalanches larger than 0.7G. The
absence of large avalanches(also reported in \cite{field}) is statistically
significant and is yet to be explained. We also studied the duration of
avalanches and found that their distribution follows a power-law ($\tau $=-3)
on a limited range covering less than one decade. The longest avalanche lasted 
26 ms.

\begin{figure}[tbp]
\end{figure}

\begin{figure}[tbp]
\epsfxsize=8.5cm
 $$\epsffile{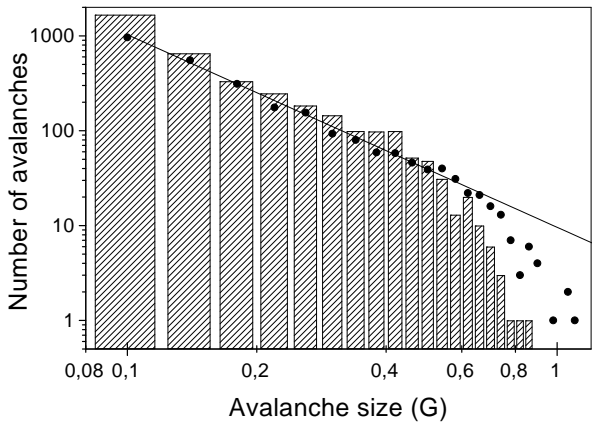}$$
\caption{Histogram of avalanches detected by a single probe (solid circles)
and those detected in the response of the entire array (columns) during
field ramps identical to the one presented in Fig. 2. The straight line
represents a power law with an exponent of -2.1. }
\label{Fig. 3}
\end{figure}

To what extent avalanches detected by a single probe are local events? In
order to answer to this question, we compared them with the response of an
entire row of Hall probes during identical ramps. Columns in Fig. 3
represent the occurrence of avalanches found in the output of an entire row
of sensors.  An avalanche of size s 
confined to an area covered by only n probes would produce a jump equal  to  ns/8  
in the output of the entire
row.  Therefore, only global avalanches would be detected with their correct
magnitude in this output.  As seen in the figure, the size collapse of the jumps detected by an 
entire row closely follows the distribution of those found by a single captor. 
However, there is a deficit of large and an excess of small avalanches in the 
response of the entire row which gives an indication of limited spatial 
extension. A direct way to explore the spatial extention of these avalanches
is to search for correlations in the events recorded by different probes. We 
found that the occurrence of an avalanche at one area of the sample makes it 
highly likely to detect a quasi-simultaneous avalanche on another place 
50 $\mu$m away. Moreover,
there is a detectable finite delay between the two detections of the same
incident. This is seen in Fig. 4 which compares the output of the first two
captors. The inset displays the difference in the amplitude of the jumps
with the shift in their temporal occurrence. The right-left asymmetry of the
data indicates that avalanches detected by captor B which is farther to the
edge occur with an average delay of the about 0.8 ms after those occurring
quasi-simultaneously at the site of captor A closer to the edge. Thus, as
one would naively expect, the apparition of avalanches at a ``higher
altitude'' on the vortexpile often precedes their passage ``downward''.
Assuming a diffusive regime, the diffusion coefficient\ for avalanches can
therefore be estimated to be\ 3 10$^{-6}m^{2}s^{-1}$. The smaller
up-down asymmetry detected in the inset of Fig. 3 suggests that avalanches
often grow in size when passing from captor A ``downward'' to captor B.

\begin{figure}[tbp]
\epsfxsize=8.5cm
 $$\epsffile{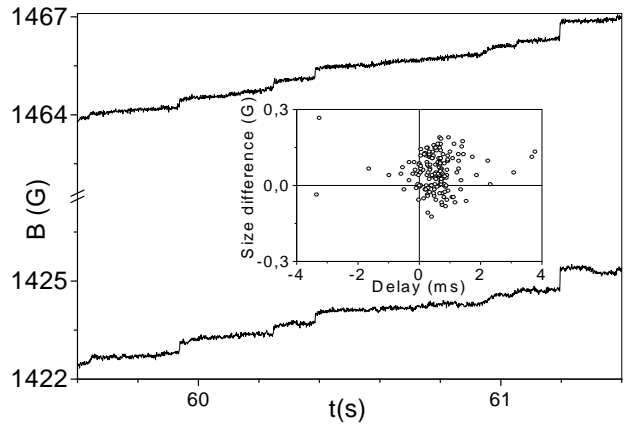}$$
\caption{Main frame: Comparison of magnetic induction recorded by two
captors 50 $\protect\mu $m apart. Most avalanches occur quasi-simultaneously
at both points. Inset: Each symbol represents an event detected
quasi-simultaneously on both probes. Difference in size between the two
events is plotted versus the difference in temporal occurence. The
right-left asymmetry of the distribution indicates that observation of an
event on the captor situated closer to the edge usually preceded its
detection on the other one.}
\label{Fig. 4}
\end{figure}

Finally, we found that below 3.4 K and for fields close to the lower
critical field, our sample displayed a quite different behavior. Huge
``catastrophic'' avalanches associated with sudden movement of thousands of
vortices were observed (see the upper inset in Fig. 5). To rule out a
thermal origin for them, we checked that the existence of these avalanches
did not depend on the speed of the field ramp. They presented a
characteristic size and were observed in both upward and downward sweeps but
with different threshold fields. They share\ all these features with vortex
avalanches observed at very low temperatures in YBa$_{2}$Cu$_{3}$CO$_{7}$ by
Zieve et al.\cite{zieve} which remain unexplained.

\begin{figure}[tbp]
\epsfxsize=8.5cm
 $$\epsffile{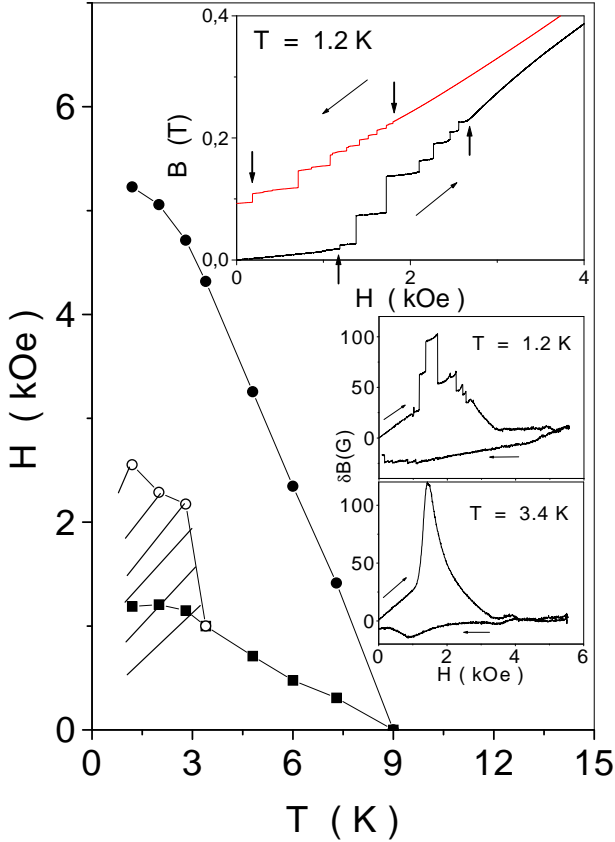}$$
\caption{Main frame: Temperature dependence of upper (filled circles) and
lower (filled squares) critical fields of the niobium sample. Catastrophic
avalanches were seen only in the marked area below 3K. Upper inset shows the
local magnetic induction at T=1.2 K as a function of increasing (decreasing)
external field with catastrophic avalanches observed in the field window
marked by upward (downward) arrows. Lower inset, compares the gradient 
in magnetic induction as a function of external field for two different 
temperatures.}
\label{Fig. 5}
\end{figure}

It is instructive to compare the gradient in magnetic induction,  
in the two different regimes. This gradient --- i.e. the critical
current density --- is readily available as $\delta$B, the difference between 
the responses of two adjacent probes. As seen in the inset, at T= 3.4 K, 
$\delta$B, after attaining a maximum  just above H$_{c1}$
decreases monotonously with increasing magnetic field. In contrast, at 
T=1.2 K and in the field range of catastrophic avalanches, it does not display
a smooth behavior suggesting  the absence of a well-defined single critical 
current. Our results suggest that in presence of strong pinning, when vortices 
are far apart and densities and thermal activation is weak, the system does not
build up a critical state. This situation may be compared with real
sandpiles where the occurrence of huge quasi-periodic avalanches was attributed
to the absence of a single ``critical'' slope\cite{nagel}. The presence of this
type of avalanches in a sample which displays a wide distribution of
avalanche scales at higher temperatures is enlightening. It indicates that
the relevant parameter in the passage between the two type of behaviors in
not the density of pinning centers as suggested by recent theoretical
suggestions\cite{bassler,pla}. Hence, the catastrophic avalanches reported
here and in ref. \cite{zieve} still lack a satisfactory explanation. We note 
that in a  model proposed by Gil and Sornette\cite{gil}, both 
large-avalanche and SOC behaviors may emerge in the same system
according to the outcome of the competition between diffusive relaxation and
instability growth rates.

In conclusion, we determined the size distribution of internal avalanches of
vortices during a slow ramp of external magnetic field. A single sample of
niobium in the strong pinning limit displayed two different types of
behavior. At temperatures above 3 K, avalanches of different sizes were
observed and the system is constantly kept in a critical state. In contrast, at
low temperatures and low fields, the occurence of huge avalanches with
characteristic size is associated with the absence of criticality.

\end{document}